\documentclass[seceqn,10pt]{article}

\usepackage{seceqn}
\usepackage{epsf}
\usepackage{cite}
\usepackage{amsmath}
\usepackage{amsfonts}
\usepackage{amssymb}
\usepackage{bm}
\usepackage{bbm}
\usepackage{graphicx}
\usepackage{epsfig}
\usepackage{latexsym}
\usepackage{nicefrac}
\usepackage{dcolumn}
\usepackage{color}

\newcolumntype{.}{D{x}{}{7}}

\usepackage{fancyhdr}
\def\corresponds{{\lower.1.377ex\hbox{=}}{\rm\kern-.75em^\triangle}}
\def\succsim{\succ\kern-.9em_\sim\kern.3em}
\def\precsim{\prec\kern-1em_\sim\kern.3em}
\def\slantfrac#1#2{\kern1em^{#1}\kern-.3em/\kern-.1em_{#2}}
\def\lfrac#1#2{{}^{#1\!}\kern-.0em/_{#2}}

\def\buildrel#1\under#2{\mathrel{\mathop{\kern0pt #2}\limits_{#1}}}

\definecolor{light}{gray}{0.90}
\definecolor{darker}{gray}{0.50}
\definecolor{dark}{gray}{0.30}

\def\dd{{\mathrm{d}}}

\def\ee{{\mathrm{e}}}

\def\calO{{\mathcal{O}}}

\def\tfrac#1#2{ {\textstyle{\frac{#1}{#2}} } }

\begin{document}

\pagestyle{empty}

\newpage

\vspace*{-2.0cm}
\begin{center}
\begin{tabular}{c}
\hline
\rule[-3mm]{0mm}{12mm}
{\large \sf Lamb Shift in Muonic Hydrogen. ---I.}\\
\rule[-5mm]{0mm}{12mm}
{\large \sf Verification and Update of Theoretical Predictions}\\
\hline
\end{tabular}
\end{center}
\vspace{0.0cm}
\begin{center}
U. D. Jentschura\\
\scriptsize
{\em Department of Physics,
Missouri University of Science and Technology,
Rolla, Missouri, MO65409, USA, and\\
National Institute of Standards and Technology,
Gaithersburg, Maryland, MD20899, USA}
\end{center}
\begin{center}
\begin{minipage}{15.0cm}
{\underline{Abstract}}
In view of the recently observed discrepancy of theory and experiment for
muonic hydrogen [R.~Pohl {\em et al.}, Nature vol.~466, p.~213 (2010)], 
we reexamine the theory on which the quantum electrodynamic (QED)
predictions are based. In particular, 
we update the theory of the $2P$--$2S$ Lamb
shift, by calculating the self-energy of the bound muon in
the full Coulomb$+$vacuum polarization (Uehling) potential. We also 
investigate the relativistic two-body corrections  
to the vacuum polarization shift, and we analyze the 
influence of the shape of the nuclear charge distribution 
on the proton radius determination. The
uncertainty associated with the 
third Zemach moment $\left< r^3 \right>_2$ in the determination of the proton
radius from the measurement is estimated. 
An updated theoretical prediction for the 
$2S$--$2P$ transition is given.
\end{minipage}
\end{center}

\noindent

Relativistic and quantum electrodynamic (QED) effects in atoms, molecules, and ions;\\
Quantum electrodynamics;\\
Specific calculations;\\
\vfill
\tableofcontents
\vfill
\begin{center}
\begin{minipage}{15cm}
\begin{center}
\hrule
{\bf \scriptsize
\noindent electronic mail: ulj@mst.edu}
\end{center}
\end{minipage}
\end{center}

\newpage

\pagestyle{fancy}
\sloppy

%
%
\section{Introduction}
\label{sec1}

The high accuracy of quantum electrodynamic (QED)
predictions and the precise spectroscopy of simple atomic systems allow for the
determination of fundamental constants like the Rydberg constant $R_\infty$
from the hydrogen spectrum, $\alpha$ from the helium fine structure, and the
electron mass $m_e$ from the $g$ factor of hydrogenlike ions.  In all these
cases, nuclear structure effects are small or can be eliminated.  On the other
hand, from a comparison of the theoretical isotope shift of optical transitions
with experimental values for such atoms as He, Li, and Be$^+$, the nuclear
charge radii of the heavy, unstable isotopes, in comparison to the stable ones,
have been determined with high accuracy~\cite{WaEtAL2004,SaEtAL2006}.  The most
elaborate measurement~\cite{PaEtAl2010} of the hydrogen-deuterium (H--D)
isotope shift of $1S$--$2S$ transition gives a deuterium-proton mean square
charge radius difference accurate to 2 parts in $10^5$.  In some cases, the
excitation of nuclei by the orbiting electron, the so called nuclear
polarizability correction, is a significant effect and has to be included in
the theoretical predictions. For example, the $1S$--$2S$ transition in D is
affected by about $20\,{\rm{kHz}}$ due to the deuteron
polarizability~\cite{FrPa1997a}, while the current experimental precision for
the H-D isotope shift is below $100\,{\rm{Hz}}$~(see
Ref.~\cite{PaEtAl2010}).

Bound muons penetrate the electron charge cloud and undergo transitions close
to the atomic nucleus, partially screened by the remaining
electrons. The determination of absolute charge radii from the 
muonic transitions is an established technique for the 
determination of nuclear radii~\cite{An2004,AnEtAl2009}.
Also, the CODATA values~\cite{MoTaNe2008} for the 
proton and deuteron nuclear radii are mainly determined by
an analysis of the 23~most accurately measured transitions 
in these systems~\cite{JeKoLBMoTa2005}.
Both atomic hydrogen as well as bound muonic systems
are essentially two-body bound systems, with a
comparatively light orbiting particle (electron or muon) and a heavier nucleus.
Still, the combined evaluation of relativistic, QED and two-body 
effects remains difficult, and the presently
achieved accuracy is the result of decade-long work of a number of physicists.
Since the original calculations of Lamb shift by Bethe, later by Feynman and
others, theory has been worked out to a very high precision. 

For hydrogen, present limitations come from the inaccuracy in the two-loop
electron self-energy and amount to about $1\,{\rm kHz}$ for the 
$1S$ Lamb shift.
This constitutes a relatively small uncertainty in comparison to the 
$1\,{\rm MHz}$ shift due to the proton finite size $r_p$ and allows for a determination
of $r_p$ with an accuracy of 8~parts per thousand, namely 
\begin{equation}
\label{rpCODATA}
r_p = 0.8768(69)\,{\rm{fm}} \qquad 
\mbox{(Refs.~\cite{JeKoLBMoTa2005,MoTaNe2008})} \,.
\end{equation}
This value is in excellent agreement with the recently obtained 
proton radius from electron scattering~\cite{BeEtAl2010}, 
which yields a value of $r_p = 0.879(8) \, {\rm fm}$.
Here, we have added the statistical and systematic uncertainties given in
Ref.~\cite{BeEtAl2010} quadratically.
The muonic hydrogen value 
\begin{equation}
\label{rpmuH}
r_p = 0.84184(67) \, {\rm fm}  \qquad
\mbox{(Refs.~\cite{PoEtAl2010})} 
\end{equation}
is in significant ($5.0\,\sigma$) disagreement with the 
CODATA and ($4.6\,\sigma$)  with the electron scattering value.
This proton radius significantly disagrees with 
a number of investigations, including determinations
of the proton radius based on previous scattering experiments
(``world scattering data'') reported in 
Refs.~\cite{Ro2000proton,Si2003proton,Si2007}.

For atomic hydrogen and more generally for electronic bound systems,
the energy shift due to the finite nuclear size is
proportional to the mean square charge radius $\langle r^2\rangle$, and the
contribution from higher moments such as the convoluted, third-order Zemach
moment $\left< r^3 \right>_2$ is negligibly small~\cite{FrSi2005}, only about
$50\,{\rm Hz}$, and similarly, proton polarizability effects amount to only
about $100\,{\rm{Hz}}$ ~(see Refs.~\cite{KhSe1998,Ro1999,KhSe2000}).
Both of the latter effects thus have only a negligible effect
on the proton radius determination from electronic hydrogen spectroscopy
(we sometimes refer to atomic hydrogen as ``electronic hydrogen''
in the current work in order to uniquely distinguish the system 
from muonic hydrogen).

Indeed, the situation is different in muonic hydrogen.  The muon Bohr radius is about
200 times smaller, and the energy levels are significantly affected by the
proton finite size. The contribution from higher order moments of the nuclear
charge distribution and from the proton polarizability, although still small,
have to be accurately estimated in order to obtain a reliable proton charge
radius. If one refrains from determining a new proton radius from
the recent measurement~\cite{PoEtAl2010} of the $2S \, (F\!\!=\!\! 1)
\Leftrightarrow 2P_{3/2} (F\!\!=\!\!2)$ transition and 
uses the CODATA proton radius in order to predict the 
transition energy, then one finds that the 
experimental result reported in Ref.~\cite{PoEtAl2010}  deviates from QED theory by
about $0.31\,{\rm{meV}}$.

A reinvestigation of the theory of the $2S$--$2P$ Lamb shift transition in
muonic hydrogen is thus indicated, especially because the 
theory of muonic atoms is not free of surprises~\cite{BrMo1978}. We proceed as follows.
In Sec.~\ref{sec2}, we first discuss the size of the observed
discrepancy in relation to the relativistic and QED effects 
which describe the energy levels of muonic hydrogen. 
We then continue to verify the current status of theory
on the basis of the two-body Breit Hamiltonian and 
evaluate relativistic effects and
quantum electrodynamic corrections (vacuum polarization and 
self energy, and combined effects) which are
relevant on the level of the current disagreement.
We then continue to calculate a few hitherto 
neglected corrections to theory (Sec.~\ref{sec3}).
Special attention is devoted to the calculation of the Bethe
logarithm in the full Coulomb$+$Uehling potential, 
which amounts to a nonperturbative treatment of the 
vacuum polarization correction to the self energy.
A final update of theory and a reevaluation of the 
proton radius is presented in Sec.~\ref{sec4}.
Conclusions are reserved for Sec.~\ref{conclu}.

%
%
\section{Verification of Theory}
\label{sec2}

%
%
\subsection{Size of the Discrepancy}

In view the recently observed discrepancy of theory and experiment in muonic
hydrogen~\cite{PoEtAl2010}, we attempt to verify the
theory~\cite{Pa1995,Pa1999,Pa2007,Bo2005mup,Bo2005mud,FaMa2003,Ma2005mup,Ma2007mup,Ma2008}
used for the theoretical evaluation of the recent
measurement~\cite{PoEtAl2010}.  
We base our considerations on the extensive
list of corrections presented in Tables~1 and~2 of the supplementary
information included with the recent paper~\cite{PoEtAl2010}, whose entries in
turn are based on previous theoretical
work~\cite{Pa1995,Pa1999,Pa2007,Bo2005mup,Bo2005mud,FaMa2003,Ma2005mup,Ma2007mup,Ma2008}.

The authors of Ref.~\cite{PoEtAl2010} report on a measurement of the
energy interval 
\begin{equation}
\label{deltaEdef}
\Delta E = E\left( 2P_{3/2}(F\!=\!2) \right) -
E\left( 2S_{1/2}(F\!=\!1) \right) 
\end{equation}
in muonic hydrogen.
Summing up all known effects, the authors
of Ref.~\cite{PoEtAl2010} use the following theoretical prediction
\begin{equation}
\label{Eth_old}
\overline E_{\mathrm{th}} = 
\left( 209.9779(49) - 5.2262 \frac{r_p^2}{\mathrm{fm}^2}  + 
0.0347 \frac{r_p^3}{\mathrm{fm}^3} \right) 
{\rm meV}.
\end{equation}
The experimental result is
\begin{equation}
\label{Eexp}
E_{\mathrm{exp}} = 206.2949(32) \, {\rm meV} \,.
\end{equation}
One thus infers the root-mean-square 
proton radius given in Eq.~\eqref{rpmuH}.

On the other hand, one can also interpret the 
result of the measurement~\cite{PoEtAl2010} as 
a disagreement of theory and experiment in an important
quantum electrodynamic measurement. Namely, if
one uses the CODATA recommended proton radius given in 
Eq.~\eqref{rpCODATA} and evaluates theory according to 
Eq.~\eqref{Eth_old}, then one obtains a theoretical 
prediction of
$E_{\rm{th}} = 205.984(63) \, {\rm meV}$
which deviates from the experimental result
given in Eq.~\eqref{Eexp} by $5.0 \, \sigma$.
As already mentioned, an additional, hypothetical
effect that would shift theory by (roughly) $+0.31\, {\rm{meV}}$ 
might thus explain the discrepancy of the 
proton radius given in Eq.~\eqref{rpmuH} and the 2006 CODATA value
of $r_p = 0.8768(69) \, {\rm{fm}}$.

The $2S_{1/2}(F\!=\!1) \Leftrightarrow 2P_{3/2}(F\!=\!2)$ 
transition frequency is the sum of three
contributions:
\begin{itemize} 
\item (i) the $2S$--$2P_{1/2}$ Lamb shift,
\item (ii) the $2P_{1/2}$--$2P_{3/2}$ fine-structure splitting, and 
\item (iii) the $2S$ and $2P_{3/2}$ hyperfine structure.
\end{itemize} 
On the level of the theoretical-experimental discrepancy
($0.31 \, {\rm meV}$), these contributions can 
be broken down into a set of well-established,
conceptually simple relativistic and QED corrections,
as detailed below.

%
%
\subsection{Reduced Mass Dependence}

Our goal is to verify the theory of the
$2S \, (F\!\!=\!\! 1) \Leftrightarrow 2P_{3/2} (F\!\!=\!\!2)$ 
transition energy in muonic hydrogen
on the level of the 
reported discrepancy~\cite{PoEtAl2010}, i.e.,
$0.31\,{\rm meV}$ in energy units.
The unperturbed nonrelativistic 
Schr\"{o}dinger Hamiltonian of the bound muon-proton 
system is 
\begin{equation}
\label{H0}
H_0 = \frac{\vec p^{\,2}}{2\,m_R}-\frac{\alpha}{r} \,,
\end{equation}
and the corresponding 
unperturbed binding energies of bound states in muonic 
hydrogen are given by
\begin{equation}
\label{E0}
E_0 = - \frac{\alpha^2 \, m_R}{2 \, n^2} \,,
\end{equation}
where $m_R = m_\mu \, m_N/(m_\mu + m_N)$ is the 
reduced mass of the system (we use units with 
$\hbar = c = \epsilon_0 = 1$).
Here, $n$ is the principal quantum number. 
The proportionality to the reduced mass
(not to the muon mass) follows from an elementary 
separation of the two-body Hamiltonian describing the 
proton and muon into relative and center-of-mass 
coordinates~\cite{LaLi1979Vol3}.
Relativistic corrections enter at relative order $\alpha^2$,
i.e., at order $\alpha^4 \, m_R$.
Their reduced mass dependence is captured in the 
two-body Breit Hamiltonian~\cite{BeLiPi1982vol4}.
For muonic hydrogen, the mass ratio of 
orbiting particle and nucleus is 
\begin{equation}
\frac{m_\mu}{m_p} = 
\frac{1}{8.880\,243\,\dots} \approx \frac{1}{9} \,,
\end{equation}
and therefore not necessarily small compared to unity.
The one-particle Dirac equation, which is 
valid only in the limit $m_\mu/m_p \to 0$, 
therefore cannot be 
used as a good approximation for the 
calculations of the relativistic effects.

Vacuum polarization effects shift the 
muonic hydrogen spectrum at relative order $\alpha$,
i.e., at order $\alpha^3 \, m_R$.
Although the vacuum polarization corrections 
are thus larger than the relativistic corrections, 
we start our discussion with the latter effects as
they constitute the most natural extension of nonrelativistic 
atomic theory without field quantization.

%
%
\subsection{Relativistic Effects and Fine Structure}

The two-body Breit-Pauli Hamiltonian for the 
muon-proton system,
including the reduced mass corrections but without
the anomalous magnetic moment of the muon, reads as follows,
\begin{align}
\label{HBP}
H_{\rm BP} =& \; -\frac{p^4}{8\,m_\mu^3}-\frac{p^4}{8\,m_p^3}
-\frac{\alpha}{2\,m_\mu\,m_p}\,
p^i\biggl(\frac{\delta^{ij}}{r}+\frac{r^i\,r^j}{r^3}\biggr)\,p^j
\nonumber \\
& \;\;
+\frac{2\,\alpha}{3}\,\biggl(\langle r^2_p\rangle
+\frac{3}{4\,m_p^2}+\frac{3}{4\,m_\mu^2}\biggr)\,\pi\,\delta^3(r)
+\frac{\alpha\,g_p}{3\,m_\mu\,m_p}\,
\vec\sigma_p\cdot\vec\sigma_\mu\,\pi\,\delta^3(r)
\nonumber\\
& \;\;
+\frac{\alpha\,g_p}{8\,m_\mu\,m_p}\,\frac{\sigma_\mu^i\,\sigma_p^j}{r^3}\,
\biggl(3\frac{r^i\,r^j}{r^2} - \delta^{ij}\biggr)
+\frac{\alpha}{4\,r^3}\,\biggl(
\frac{\vec\sigma_\mu}{m_\mu^2} +
\frac{2\,\vec\sigma_\mu+g_p\,\vec\sigma_p}{m_\mu\,m_p}
-\frac{(g_p-1)\,\vec\sigma_p}{m_p^2}
\biggr) 
\cdot\vec r\times\vec p \,.
\end{align}
The proton $g$ factor is 
$g_p = 5.585\,694\,712(46)$ (see Ref.~\cite{MoTaNe2008}).
The anomalous magnetic moment of the proton comes from
its internal structure. In the leading QED approximation,
one assumes a point proton, minimally coupled to the electromagnetic field
with its physical value of the magnetic moment which determines the
hyperfine splitting. Strictly speaking, this phenomenologically
inserted intrinsic proton magnetic moment leads to a QED theory which is not
renormalizable. The situation is different for the 
anomalous magnetic moment of the muon, which is 
described by vertex corrections within QED theory.
Lacking a convenient way to describe {\em ab initio} the inner structure of the 
proton with its anomalous magnetic moment, the phenomenological insertion
of $g_p$ into the proton-spin dependent terms in the 
Breit--Pauli Hamiltonian 
is the most obvious treatment in theoretical calculations adopted so far
in the literature.

As already mentioned above, the $2S_{1/2}(F\!=\!1) \Leftrightarrow 2P_{3/2}(F\!=\!2)$
transition frequency is the sum of three
contributions:
(i) the $2S_{1/2}$--$2P_{1/2}$ Lamb shift,
(ii) the $2P_{1/2}$--$2P_{3/2}$ fine-structure splitting, and
(iii) the $2S_{1/2}$ and $2P_{3/2}$ hyperfine structure.
The $2S_{1/2}$ and $2P_{3/2}$ hyperfine structure splittings,
as well as the $2P_{1/2}$--$2P_{3/2}$ fine-structure splitting
can directly be determined from the 
Breit--Pauli Hamiltonian given in Eq.~\eqref{HBP}.
On the level of the theoretical-experimental discrepancy
($0.31 \, {\rm meV}$), we find that the neglected terms beyond
the Breit--Pauli approximation are already too small
in magnitude to explain the discrepancy on the basis
of either the fine-structure or hyperfine-structure intervals.

We start with the $2S$ hyperfine structure, which is 
exclusively given by the terms proportional to 
$\vec\sigma_p \cdot \vec \sigma_\mu \, \delta^3(r)$ 
in the Breit--Pauli Hamiltonian.
It is sufficient to evaluate matrix elements of this 
operator, with nonrelativistic spinor wave functions 
that are solutions of the Schr\"{o}dinger Hamiltonian 
of the system (with reduced mass $m_R$).
The result is 
\begin{align}
E^{\rm BP}_{\rm{hfs}}(2S) =& \;
\langle 2S(F\!=\!1) | H_{\rm BP} | 2S(F\!=\!1) \rangle 
- \langle 2S(F\!=\!0) | H_{\rm BP} | 2S(F\!=\!0) \rangle
\nonumber\\[2ex]
\approx & \;
\frac{g_p}{6} \, \alpha^4 \, 
\frac{m_R^3}{m_\mu \, m_p} = 22.805 \, {\rm{meV}} \,.
\end{align}
After a consideration of the proton structure
(Zemach), and QED radiative corrections, 
one obtains (see Ref.~\cite{Ma2005mup}) a theoretical 
prediction of 
\begin{equation}
E^{\rm QED}_{\rm{hfs}}(2S) = 22.8148(78) \, {\rm{meV}} \,.
\end{equation}
For the $2S(F\!=\!1)$ sublevel, the difference 
\begin{equation}
\frac14 \, \left( E^{\rm QED}_{\rm{hfs}}(2S) - 
E^{\rm BP}_{\rm{hfs}}(2S) \right) = 
0.0023 \, {\rm{meV}} 
\end{equation}
already is two orders of magnitude smaller than the experimental-theoretical 
discrepancy of $0.31 \, {\rm{meV}}$. 

The $2P_{1/2}$ hyperfine splitting (energy difference of the 
$F=1$ and $F=0$ states) is not needed for the 
analysis of the experiment of Ref.~\cite{PoEtAl2010}.
Nevertheless, it is interesting to note that 
from Eq.~\eqref{HBP}, we have
\begin{align}
E^{\rm BP}_{\rm{hfs}}(2P_{1/2}) =& \;
\langle 2P_{1/2}(F\!=\!1) | H_{\rm BP} | 2P_{1/2}(F\!=\!1) \rangle 
- \langle 2P_{1/2}(F\!=\!0) | H_{\rm BP} | 2P_{1/2}(F\!=\!0) \rangle 
\nonumber\\[2ex]
= & \; \frac{g_p}{18} \, 
\frac{\alpha^4 m_R^3}{m_\mu \, m_p} \,
\left( 1 + \frac{g_p - 1}{2 \, g_p} \, 
\frac{m_\mu}{m_p} \right) 
= 7.953\, {\rm{meV}} \,.
\end{align}
Including the anomalous magnetic moment of the muon, 
one obtains a result of~\cite{Pa1995}
\begin{align}
E^{\rm QED}_{\rm{hfs}}(2P_{1/2}) = 7.963\, {\rm{meV}} \,,
\end{align}
which differs from 
$E^{\rm BP}_{\rm{hfs}}(2P_{1/2})$ 
only on the level of $0.010\, {\rm{meV}}$.

The $2P_{3/2}$ hyperfine splitting (energy difference of the
$F=2$ and $F=1$ states) is obtained from the Breit--Pauli
Hamiltonian as 
\begin{align}
E^{\rm BP}_{\rm{hfs}}(2P_{3/2}) =& \;
\langle 2P_{3/2}(F\!=\!2) | H_{\rm BP} | 2P_{3/2}(F\!=\!2) \rangle 
- \langle 2P_{3/2}(F\!=\!1) | H_{\rm BP} | 2P_{3/2}(F\!=\!1) \rangle 
\nonumber\\[2ex]
=& \; \frac{g_p}{45} \,
\frac{\alpha^4 m_R^3}{m_\mu \, m_p} \,
\left( 1 + \frac54 \frac{g_p - 1}{g_p} \, 
\frac{m_\mu}{m_p} \right) 
= 3.392\, {\rm{meV}} \,.
\end{align}
The full QED result is minimally displaced by less than 
$0.001\,{\rm meV}$ (see Ref.~\cite{Ma2008}),
\begin{align}
E^{\rm QED}_{\rm{hfs}}(2P_{3/2}) = 3.3926\, {\rm{meV}} \,,
\end{align}
The Breit--Pauli Hamiltonian also leads to an 
off-diagonal coupling of the $2P_{1/2}(F\!=\!1)$ and
$2P_{3/2}(F\!=\!1)$ sublevels~\cite{Pa1995}, 
which evaluates to 
\begin{align}
V^{\rm BP}(2P) =& \;
\langle 2P_{1/2}(F\!=\!1) | H_{\rm BP} | 2P_{3/2}(F\!=\!1) \rangle 
\nonumber\\[2ex]
=& \; -\frac{g_p}{144\,\sqrt{2}} 
\frac{\alpha^4 m_R^3}{m_\mu \, m_p} 
\left( 1 + 2 \frac{g_p - 1}{g_p} 
\frac{m_\mu}{m_p} \right) 
= -0.796 \, {\rm{meV}},
\nonumber
\end{align}
in excellent agreement with the value given in Eq.~(85)
of Ref.~\cite{Pa1995}. 

Finally, the energy difference of the hyperfine 
centroids (the fine-structure splitting) of the $2P$ states
is readily evaluated as
\begin{align}
E^{\rm BP}_{\rm{fs}}(2P) =& \;
\langle 2P_{3/2} | H_{\rm BP} | 2P_{3/2} \rangle -
\langle 2P_{1/2} | H_{\rm BP} | 2P_{1/2} \rangle 
= \frac{\alpha^4 m_R^3}{32 \, m_\mu \, m_p} \,
\left( 1 + 2 \frac{m_\mu}{m_p} \right) 
= 8.329\, {\rm{meV}} \,.
\nonumber
\end{align}
This result differs by only $0.023\,{\rm meV}$ from the full 
QED result of~\cite{Pa1995,Bo2005mup}.
\begin{equation}
E^{\rm QED}_{\rm{fs}}(2P) = 8.352\,{\rm meV} \,.
\end{equation}
If we define the states 
\begin{align}
| 1 \rangle =& \; | 2P_{1/2} (F\!=\!0) \rangle \,,\qquad
| 2 \rangle = | 2P_{1/2} (F\!=\!1) \rangle \,, \qquad
| 3 \rangle = | 2P_{3/2} (F\!=\!1) \rangle \,, \qquad
| 4 \rangle = | 2P_{3/2} (F\!=\!2) \rangle \,,
\end{align}
and the matrix elements
\begin{align}
\beta_{1/2} =& \; E_{\rm{hfs}}(2P_{1/2})  \,, \qquad
v = V(2P) \,, \qquad
\beta_{3/2} = E_{\rm{hfs}}(2P_{3/2})  \,, \qquad
f = E_{\rm{hfs}}(2P) \,,
\end{align}
and the zero point of the energy scale to be the hyperfine centroid 
of the $2P_{1/2}$ levels, then the Breit--Pauli Hamiltonian
in the $2P$ state manifold assumes the following matrix form $M_{\rm BP}$:
\begin{equation}
\label{MBP}
M = \left( \begin{array}{cccc} 
-\tfrac34 \, \beta_{1/2} & 0 & 0 & 0 \\[2ex]
0 & \tfrac14 \, \beta_{1/2} & v & 0 \\[2ex]
0 & v & -\tfrac58 \, \beta_{3/2} + f & 0 \\[2ex]
0 & 0 & 0 & \tfrac38 \, \beta_{3/2} + f \\[2ex]
\end{array}
\right) \,.
\end{equation}
The off-diagonal elements $v$ lead to admixtures 
to the $|2P_{1/2}(F\!=\!1)\rangle$ levels from the 
$|2P_{3/2}(F\!=\!1)\rangle$ levels and vice versa,
and to a repulsive interaction as for any coupled
two-level system. In agreement with this 
general consideration,
a diagonalization of $M_{\rm BP}$ immediately leads to 
the conclusion, that the $|2P_{1/2}(F\!=\!1)\rangle$
is lowered in energy by 
\begin{equation} 
\Delta = 0.145 \, {\rm meV} \,,
\end{equation}
whereas the $|2P_{3/2}(F\!=\!1)\rangle$ energy 
is increased by $\Delta$. This is in full agreement
with Ref.~\cite{Pa1995}.

In the current derivation, we have not distinguished 
the magnetic projections of the hyperfine sublevels.
If these are included, a 12-dimensional matrix 
is obtained from the singlet $| 1\rangle$, the
three magnetic sublevels of states $|2\rangle$ and $|3\rangle$, and the 
five magnetic sublevels of state $|4\rangle$.
However, we have checked by an explicit calculation
of the Hamiltonian matrix, using angular momentum 
algebra~\cite{Ed1957,VaMoKh1988}, that the energies 
obtained by diagonalizing the full Hamiltonian 
matrix are the same as those obtained from~\eqref{MBP},
and that, in particular, the state $|2P_{3/2}(F\!=\!2) \rangle$,
which is so important for the experiment~\cite{PoEtAl2010},
remains uncoupled from the other hyperfine and 
fine-structure levels.

%
%
\subsection{QED and Lamb Shift}

The theory of the $2P_{1/2}$--$2S_{1/2}$ Lamb shift in muonic
hydrogen is surprisingly simple.
One here speaks of the $2P$--$2S$ Lamb shift because the 
$2P$ level is energetically higher than $2S$, in contrast 
to electronic hydrogen where the situation is opposite,
and the $2S$--$2P$ Lamb shift is observed.
The one-loop vacuum-polarization (Uehling) correction
due to virtual electron-positron pairs
gives the main contribution, 
and already the first-order effect (in perturbation theory)
accounts for $99.5\,\%$ of the Lamb shift,
or $205.0074\, {\rm meV}$. The second-order 
perturbation theory effect contributes $0.1509\,{\rm meV}$,  
and the two-loop vacuum polarization 
gives a shift of $1.5081\,{\rm meV}$.
We note that
in the 1970s, an error in the evaluation of
the two-loop vacuum polarization~\cite{Fr1969} has led to a discrepancy
of theory and experiment for heavy muonic ions~\cite{DiEtAl1971}.
The discrepancy was reduced after the error 
was discovered~\cite{Bl1972,SuWa1972,Be1973}.
This observation implies that a careful verification of the 
vacuum polarization effects appears worthwhile in the 
current situation.

Finally, the one-loop self-energy of the 
bound muon decreases the $2P$--$2S$ Lamb shift
and amounts to $-0.6677\,{\rm meV}$ (the sign is natural 
as the self-energy effect is the dominant effect in 
electronic hydrogen, where the sign of the entire 
Lamb shift is reversed compared to muonic hydrogen,
due to this term). 
The four mentioned contributions 
to the Lamb shift, which are recalculated and verified below,  
already account for $205.9987\, {\rm meV}$.
The total QED result without the nuclear
finite-size effect, according to Ref.~\cite{PoEtAl2010},
is $206.0573\,{\rm meV}$.
The difference, $0.0586\,{\rm meV}$, is already 
much smaller than the observed discrepancy
of $0.31\,{\rm meV}$, and there are no cancellations 
among conceivably large, further QED effects not 
accounted for in the above list of dominant
vacuum-polarization and self-energy effects. 
This illustrates that the theory of the Lamb shift in 
muonic hydrogen rests, to a large extent, on a 
very well established subset of QED effects.

%
%
\subsection{One--Loop Vacuum Polarization}

The one-loop vacuum polarization effect is 
described by the Uehling Potential $V_{\rm vp}(r)$ 
which was evaluated already in 1935 (Ref.~\cite{Ue1935}).
Muonic hydrogen is different from 
electronic hydrogen because the classical 
orbit of the bound muon is much closer to the 
proton than for the electron.
It is thus useful to analyze the 
asymptotic behavior of the Uehling potential for large and small
distances from the nucleus. We define the scaled coordinate
$\rho = \alpha \, m_R \, r$, where $\alpha$ is the 
fine-structure constant, and $m_R$ is the reduced mass of the bound system.
The Coulomb potential scales as
\begin{equation}
\label{V}
V(r) = - \frac{\alpha}{r} 
= - \frac{\alpha^2 \, m_r}{\rho} \,.
\end{equation}
The one-loop Uehling potential is
\begin{equation}
\label{Vvp}
V_{\rm vp}(r) = - \frac{\alpha^3  m_r}{\pi \rho} \int\limits_1^\infty \dd u \,
\ee^{-2 u \, \rho \, x} 
\frac{\sqrt{u^2 - 1} \,  (2 u^2 + 1)}{3 u^4} \,,
\end{equation}
where $x = \frac{m_e}{\alpha \, m_r} = 0.73738368\ldots$ for muonic hydrogen.
In atomic hydrogen, $x \approx \alpha^{-1} \gg 1$, and the Uehling potential
can be approximated by a Dirac $\delta$, but this is not the case 
for muonic atoms. For small distances ($\rho \to 0$), one finds,
in agreement with Ref.~\cite{Bl1972},
\begin{equation}
V_{\rm vp}(r) \sim \frac{\alpha^3 m_r}{\pi\,\rho} 
\left[ \frac23 \, \biggl(\ln(\rho \, x) + \gamma_E \biggr)
- \frac{\pi}{2} \, \rho \, x + \frac59 \right] +
\calO(\rho) \,.
\end{equation}
As compared to the Coulomb potential, only a logarithmic divergence
is added. For large distances, $\rho \to \infty$, by contrast, 
the Uehling potential decreases exponentially,
\begin{equation}
V_{\rm vp}(r)  \sim - \frac{\alpha^3 \, m_r}{\sqrt{\pi}} \,
\ee^{-2 \rho x} \, 
\left[ \frac{1}{4 \, \rho^{5/2} \, x^{3/2}} 
- \frac{29}{64 \, \rho^{7/2} \, x^{5/2}}
+ \frac{2225}{2048 \, \rho^{9/2} \, x^{7/2}} +
\calO\left( \frac{1}{\rho^{11/2}} \right) \right] \,,
\end{equation}
which leads to rapidly convergent radial integrals. 
On the Bohr radius scale of muonic atoms as measured 
by the scaled coordinate $\rho$,
the Uehling potential goes as $\ln(\rho)/\rho$
for $\rho \to 0$. This is more singular than  
the Coulomb potential, but only by a logarithm.
The smooth behavior excludes conceivable nonperturbative effects 
which could be expected for a 
singular behavior near the origin.

The question then is whether one should start the evaluation of the one-loop
vacuum polarization correction from the nonrelativistic Schr\"{o}dinger wave
function or from relativistic Dirac theory. The latter approach is indicated
for highly charged ions, where relativistic effects dominate over the
reduced-mass corrections.  Muonic hydrogen, on the other hand, is a light
system, and the parameter characterizing the reduced-mass corrections
($m_\mu/m_p$) is much larger than the parameter characterizing the expansion
into Coulomb vertices ($Z\alpha = \alpha$, where $Z=1$ is the charge number of
the proton).  We therefore start from the Schr\"{o}dinger Hamiltonian and add
reduced-mass effects later on the basis of a radiatively corrected Breit--Pauli
Hamiltonian.

Up to now (Refs.~\cite{Pa1995,Bo2005mup}), 
the vacuum-polarization corrections
have been evaluated using an analytic approach, 
with analytic representations of the reduced 
Green functions~(see Ref.~\cite{Pa1995} for relevant formulas). 
Here, in order to check for any conceivable 
calculational errors in the numerically dominant 
vacuum polarization corrections, we choose a 
different approach. 
Namely, because the singularity of the Uehling potential
at the origin is only logarithmic,
the combined Coulomb$+$Uehling 
potential is eligible for a numerical solution of
the Schr\"{o}dinger equation on a lattice~\cite{SaOe1989}.
In this approach, one may choose to evaluate 
perturbative terms using unperturbed Schr\"{o}dinger 
eigenstates (Green functions are represented
by the pseudospectrum of states obtained from the 
finite lattice), or one may choose to perform 
a nonperturbative solution of the Schr\"{o}dinger equation,
for the combined Coulomb$+$Uehling potential.

Performing a matrix element calculation on the 
lattice composed of an exponential grid (see Ref.~\cite{Jo2007}) 
and observing the apparent convergence on lattices
with more than $N = 300$ grid points, 
and increasing $N$ in steps of $\Delta N = 50$, 
we confirm the results,
\begin{align}
\label{deltaE1}
\delta E^{(1)} =& \; \langle 2P | V_{\rm  vp}| 2P \rangle 
- \langle 2S | V_{\rm  vp}| 2S \rangle = 205.0074 \, {\rm meV}
\end{align}
in first order and
\begin{align}
\label{deltaE2}
& \delta E^{(2)} =
\left< 2P \left| V_{\rm vp}\,\frac{1}{(E_0-H_0)'}\,V_{\rm  vp} \right| 2P \right> 
- \left< 2S \left| V_{\rm vp}\,\frac{1}{(E_0-H_0)'}\,V_{\rm  vp} \right| 2S \right> 
= 0.1509\,{\rm meV}
\end{align}
in second order. Here $H_0$ and $E_0$ are the Schr\"{o}dinger Hamiltonian
of the system and the binding energy given in 
Eqs.~\eqref{H0} and~\eqref{E0}, and 
$[1/(E_0-H_0)']$ denotes the reduced Green function.

The Hamiltonian of muonic hydrogen, including
vacuum polarization, is
\begin{equation}
\label{H0vp}
H_{0\rm vp} = \frac{\vec p^{\,2}}{2\,m_R}-\frac{\alpha}{r}+V_{\rm vp}(r) \,,
\end{equation}
For the numerical calculation,
we add the Uehling potential to the 
Schr\"{o}dinger Hamiltonian and diagonalize the 
Hamiltonian $H_{0\rm vp}$ on a pseudospectrum of states
that is generated by a finite, 
exponential grid~\cite{SaOe1989,Jo2007}.
We use the exact reduced mass of the system,
which cannot be done consistently when using the Dirac equation,
and a variable coupling parameter 
$\chi \equiv \alpha/\pi$, for various values of $\chi$,
in order to control the convergence of the calculation.
We find a smooth dependence on $\chi$, which can be 
fitted to excellent accuracy by a power series in 
$\chi$. The first two terms are consistent with 
the results derived in Eqs.~\eqref{deltaE1} and~\eqref{deltaE2}.
For $\chi =  \alpha/\pi$, which is the physically 
relevant coupling parameter for muonic hydrogen,
our result for the total $2P$--$2S$ electronic vacuum
polarization contribution to the Lamb shift of
$205.1584(1)\,{\rm{meV}}$ is in excellent agreement
with the sum of the one-loop term ($205.0074\,{\rm{meV}}$),
the iterated one-loop term (second order, $0.1509\,{\rm{meV}}$) and the
third-order perturbation theory effect of $0.00007\,{\rm{meV}}$
as discussed in Refs.~\cite{KiNi1999,IvKoKa2007comment}.

\begin{figure}[t!]
\begin{center}
\begin{minipage}{0.7\linewidth}
\begin{center}
\includegraphics[width=0.91\linewidth]{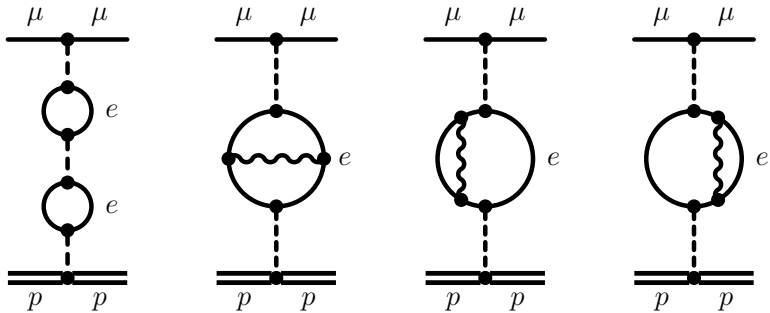}
\end{center}
\caption{\label{fig1} Two-loop vacuum-polarization diagrams.}
\end{minipage}
\end{center}
\end{figure}

%
%
\subsection{Two--Loop Vacuum Polarization}

The two-loop vacuum polarization correction
corresponds to the diagrams in Fig.~\ref{fig1}.
The expression for the diagrams was first derived 
by Kallen and Sabry in 1955 (Ref.~\cite{KaSa1955}), and the corrections
therefore carry their name.
The two-loop calculation leading to their derivation 
is non-trivial, and it is thus imperative to clarify 
the status of the two-loop potential in the literature.
A clear exposition of the derivation is not only given
in volume III of the monograph~\cite{Sc1970},
but the two-loop effect was also recalculated and later
used as input for the three-loop corrections to 
the electron $g$ factor; explicit results are
indicated in Ref.~\cite{BaRe1973}.
Finally, the result has later been generalized to non-Abelian
gauge theories, see Refs.~\cite{Kn1990,BrFl1993,ChKuSt1996}.


An evaluation of the two-loop potential with nonrelativistic Schr\"{o}dinger
wave functions, using the two-loop potential 
$V_{\rm  vp}^{(2)}$ given in Refs.~\cite{BaRe1973,Pa1993pra}
then leads to the result
\begin{align}
\label{deltaEKS}
\delta E_{\rm KS} =& \; \langle 2P | V_{\rm  vp}^{(2)} | 2P \rangle 
- \langle 2S | V_{\rm  vp}^{(2)} | 2S \rangle
= 1.5081 \, {\rm meV}\,.
\end{align}
For reference, it is quite instructive to 
divide the calculation into an evaluation of the 
loop-after-loop correction in the leftmost diagram
of Fig.~\ref{fig1} (which gives a contribution of 
$0.25\,{\rm{meV}}$ to the $2S$-$2P$ Lamb shift), 
and the ``true'' two-loop effects
diagrammatically represented in three rightmost diagrams
in Fig.~\ref{fig1}, which contribute $1.25\,{\rm{meV}}$.
The total value of $1.50\,{\rm{meV}}$ for the two-loop
effect~\cite{Pa1995,Bo2005mup} is thus confirmed.

%
%
\subsection{Muon Self--Energy}

The self-energy shifts of the $2S$ and $2P_{1/2}$ levels are
given by 
\begin{align}
\label{SE2S}
E_{\rm SE}(2S_{1/2}) =& \;
\frac{\alpha^5 \, m_R^3}{8 \, \pi \, m^2_\mu} \, 
\left\{ 
\frac43 \, \ln\left( \frac{m_\mu}{\alpha^2 \, m_R} \right)
+ \frac{10}{9}
- \frac43 \, \ln k_0(2S) 
+ 4 \pi \alpha \left( \frac{139}{128} - \frac12 \, \ln(2) \right)
\right\} \,.
\end{align}
This expression has been derived originally in the early 
days of quantum electrodynamics~\cite{LambFirst}, and the effect of 
relative order $\alpha$ has been derived in 
Refs.~\cite{Ba1951,BaBeFe1953}.
The coefficients agree with very precise numerical investigations
of the one-loop self-energy~\cite{JeMoSo1999}, 
in the domain of low nuclear charge numbers,
and an extensive list of Bethe logarithms $\ln k_0(nP)$
has been given in Ref.~\cite{JeMo2005bethe}.

There is a further effect due to virtual muon-antimuon pairs that modify the
vacuum polarization potential,
\begin{align}
\label{VP2S}
E_{\rm VP}(2S_{1/2}) =& \;
\frac{\alpha^5 \, m_R^3}{8 \, \pi \, m_\mu^2} \, 
\left\{ -\frac{4}{15} + \pi \, \alpha \frac{5}{48} \right\} \,.
\end{align}
The reduced-mass dependence of the corrections listed in 
Eqs.~\eqref{SE2S} and~\eqref{VP2S} has been analyzed in 
Ref.~\cite{SaYe1990}.

The self-energy effect on the $2P$ state is given by
\begin{align}
\label{SE2P}
E_{\rm SE}(2P_{1/2}) =& \;
\frac{\alpha^5 \, m_R^3}{8 \, \pi \, m^2_\mu} \, 
\left\{ 
- \frac{1}{6}\frac{m_\mu}{m_R}
- \frac43 \, \ln k_0(2S) \right\} \,.
\end{align}
The first term in curly brackets is due to the 
anomalous magnetic moment of the muon.
Its reduced-mass dependence, which is different 
from that of all other terms, follows from the 
Breit--Pauli Hamiltonian if one corrects the
muon-spin dependent terms by the muon anomalous magnetic moment~\cite{SaYe1990}.
Otherwise, the coefficients in Eq.~\eqref{SE2P} 
have been verified against very precise numerical
calculations~\cite{JeMoSo2001pra} of the 
one-loop self-energy. There is no 
additional vacuum-polarization effect for the 
$2P_{1/2}$ state to the order under investigation 
($\alpha^6\,m_R$).

The final result for the self-energy and muonic-vacuum
polarization shift is
\begin{align}
\label{deltaESE}
\delta E_{\rm SE} = & \;
E_{\rm SE}(2P_{1/2}) -
E_{\rm VP}(2S_{1/2}) -
E_{\rm SE}(2S_{1/2}) 
= -0.6677 \, {\rm meV} \,.
\end{align}
Finally, the contributions given in 
Eqs.~\eqref{deltaE1},~\eqref{deltaE2},~\eqref{deltaEKS}
and~\eqref{deltaESE} add up to
\begin{align}
\delta E =& \; 
\delta E^{(1)} +
\delta E^{(2)} +
\delta E_{\rm KS} +
\delta E_{\rm SE} 
= 205.9987\, {\rm meV} \,,
\end{align}
All further QED corrections listed in Table~1 of
Ref.~\cite{PoEtAl2010} (supplementary material)
add up to $0.0586\,{\rm meV}$, which gives 
a total QED result of $206.0573\,{\rm meV}$.
Both individually as well as collectively, the 
remaining effects are thus too small
to explain the observed discrepancy
of $0.31\,{\rm meV}$.

%
%
\section{Advancing Theory}
\label{sec3}

%
%
\subsection{Relativistic Effects and Vacuum Polarization}

We have been unable to resolve the discrepancy 
of theory and experiment observed in the 
recent measurement~\cite{PoEtAl2010}. Therefore,
we now turn our attention to the evaluation 
of a few numerically tiny, but still important 
corrections to the Lamb shift, which have
not yet been addressed in the literature.

The authors of
Ref.~\cite{Bo2005mup,PoEtAl2010} use a value of 
$205.0282 \,{\rm{meV}}$ for the
first-order Uehling correction 
$\delta E^{(1)}$ to the $2P$--$2S$ muonic hydrogen Lamb shift, taken
with the exact one-body Dirac wave function for both the 
$2P_{1/2}$ as well as the $2S_{1/2}$ states.
The relativistic correction thus is 
\begin{equation}
\delta E^{\rm (1b)}_{\rm rel} = 
(205.0282 -205.0074) \,{\rm{meV}} = 0.0208\,{\rm{meV}}
\end{equation}
for the one-body Dirac theory.
However, the one-body Dirac theory cannot 
account for the two-body reduced-mass corrections.
As shown in Ref.~\cite{VePa2004}, the full
two-body treatment has to be based on a generalization of Eq.~\eqref{HBP} to a
massive photon, integrated over the spectral function describing the virtual
electron-positron pairs in the vacuum polarization 
loop~\cite{BeLiPi1982vol4}.
It is imperative to use the full two-body treatment for muonic 
hydrogen because the parameter governing the relativistic
corrections captured in the one-body Dirac equation
(the fine-structure constant~$\alpha$) is much smaller for muonic hydrogen than the 
parameter characterizing the reduced-mass corrections
($m_\mu/m_p$) which are captured in the Breit--Pauli
Hamiltonian.
According to Eq.~(25) of Ref.~\cite{VePa2004}, the relativistic correction to
the one-loop electronic vacuum polarization thus reads
\begin{equation}
\delta E^{\rm (2b)}_{\rm rel} = 0.0169\,{\rm{meV}} \,,
\end{equation}
and we prefer to use this value for our final theoretical
evaluations.
The additional two-body relativistic correction beyond
the one-body treatment used in Refs.~\cite{Bo2005mup,PoEtAl2010} 
thus is 
\begin{equation}
\delta E_a = 
\delta E^{\rm (2b)}_{\rm rel} - 
\delta E^{\rm (1b)}_{\rm rel} = -0.0039 \, {\rm{meV}} \,.
\end{equation}

\begin{figure}[t!]
\begin{center}
\begin{minipage}{0.7\linewidth}
\begin{center}
\includegraphics[width=0.91\linewidth]{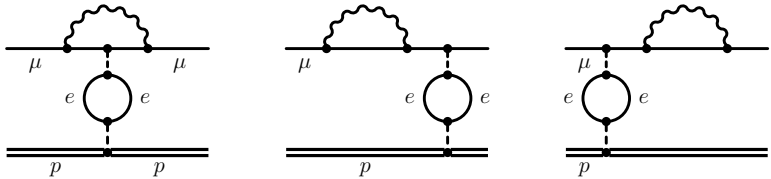}
\end{center}
\caption{\label{fig2} Electronic vacuum polarization 
corrections to the muon self-energy.}
\end{minipage}
\end{center}
\end{figure}

%
%
\subsection{Self--Energy Vacuum Polarization Corrections}

Because the dominant effect to the Lamb shift 
is due to the one-loop vacuum polarization potential
given in Eq.~\eqref{Vvp}, it would be highly desirable
to carry out the calculation of the 
one-loop self-energy effect directly for eigenstates
of the full Schr\"{o}dinger--Uehling Hamiltonian
given in Eq.~\eqref{H0vp}.
If the Uehling potential is treated in first order,
then the Feynman diagrams are given in Fig.~\ref{fig2}.

We thus consider the modification of the muon self energy due to 
the Uehling potential, which corresponds to the 
replacement $V \to V + V_{\rm vp}$ in the 
Schr\"{o}dinger Hamiltonian. 
The effect of high-energy virtual photons
in the self-energy loops given in Fig.~\ref{fig2}
can then be expressed in terms of the Dirac $F_1$ 
form factor acting on the vacuum polarization
potential $V_{\rm vp}$. 
When rewritten in terms of the noncovariant photon energy cutoff
$\epsilon$, which is a convenient overlapping parameter 
in Lamb shift calculations~\cite{Pa1993}, we have
\begin{equation}
E^{(\rm vp)}_H = \frac{\alpha}{3 \pi m_\mu^2} \, 
\left< \psi \left| \vec\nabla^2 (V+V_{\rm vp}) \right| \psi \right>
\left( \ln\left( \frac{m_r}{2 \epsilon} \right) + \frac{10}{9} \right) \,,
\end{equation}
where $| \psi \rangle$ is the wave function of the 
bound state in the full binding potential $V + V_{\rm vp}$.
Denoting by $\phi$ the Schr\"{o}dinger--Coulomb eigenstate, we have
in leading order 
\begin{equation}
\left< \phi \left| \vec\nabla^2 V \right| \phi \right> =
\frac{4 \alpha^4 m_R^3}{n^3} \; \delta_{\ell 0} \,,
\end{equation}
and thus
\begin{equation}
E^{(\rm 0)}_H = \frac{4 \alpha^5 \, m_R^3}{3 \pi m_\mu^2 \, n^3}  
\delta_{\ell 0} 
\left( \ln\left( \frac{m_\mu}{2 \epsilon} \right) + \frac{10}{9} \right) \,.
\end{equation}
Numerically, we find that the correction 
$\delta E_H $ to the high-energy part 
due to the vacuum polarization can be conveniently 
expressed in terms of a parameter $V_{61}$,
\begin{align}
\delta E_H =& \; E^{(\rm vp)}_H - E^{(\rm 0)}_H 
= \frac{\alpha^6 \, m_R^3}{\pi^2 m_\mu^2 \, n^3} \, V_{61} \,
\left\{ \ln\left( \frac{m_\mu}{2 \epsilon} \right) + \frac{10}{9} \right\} \,.
\end{align}
By a numerical diagonalization of the Schr\"{o}dinger--Uehling
Hamiltonian on an exponential grid~\cite{Jo2007}, we find
for the $V_{61}$ coefficient (the first subscript gives the 
power of $\alpha$, the second indicates the power of the logarithm),
\begin{align}
V_{61}(2S_{1/2}) =& \; 3.09 \,, \qquad
V_{61}(2P_{3/2}) = -0.023 \,.
\end{align}
The anomalous magnetic moment of the electron 
gives rise to a further combined self-energy vacuum polarization
correction,
\begin{align}
\delta E_M =& \; \frac{\alpha}{4 \pi m_\mu^2} \, 
\left< \psi \left| \frac{1}{r} \, \frac{\partial (V + V_{\rm vp})}{\partial r} 
\left( \vec\sigma \cdot \vec L \right) \right| \psi \right> 
- \frac{\alpha}{4 \pi m_\mu^2} \, 
\left< \phi \left| \frac{1}{r} \, \frac{\partial V}{\partial r} 
\left( \vec\sigma \cdot \vec L \right) \right| \phi \right> 
\nonumber\\[2ex]
=& \; \frac{\alpha^6 \, m_R^3}{\pi^2 m_\mu^2 \, n^3} \, M_{60} \,,
\end{align}
where $\psi$ again is the state in the Uehling$+$Coulomb potential,
and $\phi$ is the unperturbed state.
The matrix element $M_{60}$ is nonvanishing for $P$ states,
\begin{equation}
M_{60}(2S_{1/2}) = 0 \,, 
\qquad
M_{60}(2P_{1/2}) = -0.022 \,.
\end{equation}
The low-energy part of the muon self-energy, in the Coulomb
potential, is
\begin{equation}
E^{(0)}_L = 
\frac{4 \alpha^5 m_R^3}{3\pi m_\mu^2 \, n^3} 
\left[ \delta_{\ell 0} 
\ln\left( \frac{\epsilon}{\alpha^2 m_R} \right) - \ln k_0 \right] .
\end{equation}
It follows from second-order perturbation theory with 
time-independent field operators and an upper, noncovariant
cutoff for the photon energy~\cite{Be1947,Pa1993,JePa1996}.
We have numerically calculated the low-energy 
$\delta E^{(\rm vp)}_L$ part for solutions of the full
Schr\"{o}dinger--Uehling
Hamiltonian on an exponential grid
and find that it can be conveniently expressed as
\begin{align}
\delta E_L =& \; \delta E^{(\rm vp)}_L - \delta E^{(0)}_L
= \frac{\alpha^6 m_R^3}{\pi^2 m_\mu^2 \, n^3} \,
V_{61} \,
\ln\left( \frac{\epsilon}{\alpha^2 m_R} \right) 
- \frac{4 \alpha^6 m_R^3}{3 \pi^2 m_\mu^2 \, n^3} \,
L_{60} \,,
\end{align}
where $V_{61}$ parameterizes the modification of the 
logarithmic coefficient and equals the corresponding 
correction from the high-energy part, leading to a cancellation
of the $\epsilon$ parameter.
The coefficient $L_{60}$ gives the 
modification of the Bethe logarithm due to the 
Uehling potential. Numerically, we find
\begin{align}
L_{60}(2S_{1/2}) =& \; 11.28 \,, \qquad
L_{60}(2P_{1/2}) = -0.0146 \,.
\end{align}
In the total correction $\delta E_H + \delta E_L$, 
the auxiliary overlapping parameter
$\epsilon$ cancels, and we have
for the combined self-energy vacuum polarization 
correction,
\begin{align}
\delta E_{\rm svp} =& \;
\frac{\alpha^6 m_R^3}{\pi^2 m_\mu^2 \, n^3} \;
\left\{
\left[ \ln\left( \frac{m_\mu}{2 \alpha^2 m_R} \right) + 
\frac{10}{9} \right]
\Delta V_{61} + \Delta M_{60} - \frac43 \Delta L_{60} \right\}
\nonumber\\[2ex]
& \; = -0.0025 \, {\rm meV}\,,
\end{align}
when evaluated for the difference of the $2P$ and $2S$ states in
muonic hydrogen.
We have used the definitions,
\begin{subequations}
\begin{align}
\Delta V_{61} =& \; V_{61}(2P_{1/2}) - V_{61}(2S_{1/2}) \,,
\\[2ex]
\Delta M_{60} =& \; M_{60}(2P_{1/2}) - M_{60}(2S_{1/2}) \,,
\\[2ex]
\Delta L_{60} =& \; L_{60}(2P_{1/2}) - L_{60}(2S_{1/2}) \,.
\end{align}
\end{subequations}
Electronic vacuum polarization insertions into the 
bound muon line 
have previously  been evaluated in Eqs.~(40) and (45) of
Ref.~\cite{Pa1995} in the leading logarithmic 
accuracy as 
\begin{equation}
\delta E_{\rm svp}^{\rm log} = -0.006\, {\rm{meV}}.
\end{equation}
The modification due to our complete
treatment thus is 
\begin{equation}
\delta E_b = \delta E_{\rm svp} -
\delta E_{\rm svp}^{\rm log} = +0.0035\, {\rm{meV}}.
\end{equation}

\begin{figure}[t!]
\begin{center}
\begin{minipage}{0.7\linewidth}
\begin{center}
\includegraphics[width=1.0\linewidth]{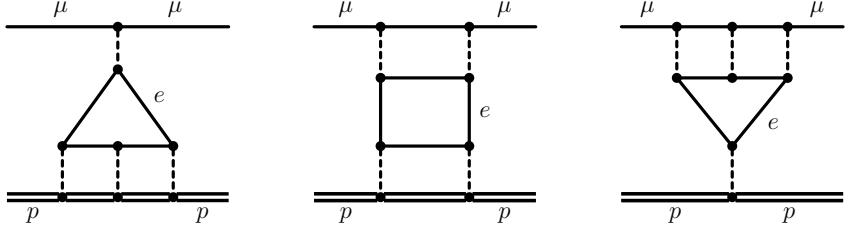}
\end{center}
\caption{\label{fig3} Light-by-light scattering diagrams.}
\end{minipage}
\end{center}
\end{figure}

%
%
\subsection{Virtual Light--by--Light Scattering}

Finally, we also include the results for the light-by-light scattering graphs
shown in Fig.~\ref{fig3}, as reported in Ref.~\cite{KaIvKoSh2010}, which add up
to 
\begin{equation}
\delta E_{\rm LL} = -0.00089(2)\,{\rm{meV}}
\end{equation}
for the $2P$--$2S$ Lamb shift.  These entries
replace the results for the virtual Delbr\"{u}ck scattering and the
Wichmann--Kroll term in~\cite{PoEtAl2010} (entries \# 9 and \# 10 in 
Table~I of the supplementary material included with Ref.~\cite{PoEtAl2010}), 
which otherwise add up to
\begin{equation}
\delta E_{\rm WK+VD} = +0.00032(135)\,{\rm{meV}}
\end{equation}
The shift is 
\begin{equation}
\delta E_c = \delta E_{\rm LL} - \delta E_{\rm WK+VD} 
= -0.0012\,{\rm{meV}} \,,
\end{equation}
and the
corresponding uncertainty of $\pm 0.00135 \,{\rm{meV}}$ is eliminated from the
theoretical prediction. 

%
%
\subsection{Zemach Moment Correction}

For $S$ states bound to an infinitely heavy nucleus, the finite nuclear-size 
(NS) effect is given by 
\begin{align}
\label{ENS}
\delta E_{\rm NS} =& \; 
= \frac{2 \alpha^4 m_R^3}{3 n^3} \delta_{\ell 0} \, \langle r^2 \rangle \, 
\left( 1 - \frac{\alpha}{2} \, \frac{m_\mu}{m_p}  \,
\frac{m_p \langle r^3 \rangle_{(2)}}{\langle r^2 \rangle} \right) \,.
\end{align}
where $\langle r^2 \rangle$ is the mean-square charge radius of the proton, 
and the $\alpha^5$ correction to the nuclear size effect
involves the so-called third Zemach moment
$\langle r^3 \rangle_{(2)}$ 
(see Ref.~\cite{Ze1956}), which results from the elastic 
part of the two-photon exchange diagrams of bound particle and nucleus, 
when the both Coulomb photon 
interactions are corrected for the finite-size effect.

The coordinate-space representation of the 
third Zemach moment reads
\begin{equation}
\langle r^3 \rangle_{(2)} =
\int \dd^3 r \; \int \dd^3 r' \; 
r^3 \, f_E(\vec r - \vec r') \; f_E(\vec r') \,,
\end{equation}
where $f_E$ is the normalized electric charge
distribution of the proton, with 
$\int \dd^3 r \, f_E^2(\vec r) = 1$.

The third Zemach moment can be obtained, in a model-independent
way, from experimentally obtained 
scattering data for electron-proton scattering,
in terms of the measured $G_E$ Sachs form factor as
\begin{equation}
\langle r^3 \rangle_{(2)} =
\frac{48}{\pi} \int_0^\infty \frac{\dd q}{q^4} \, 
\left[ G_E^2(q^2) - 1 + \frac13\, q^2 \, \left< r^2 \right> \right] \,,
\end{equation}
where the two subtraction terms ensure the 
convergence of the integral for small $q$.
From a model-independent, careful analysis of the world scattering
data performed in Ref.~\cite{FrSi2005}, Friar and Sick
obtain the value
\begin{equation}
\label{third}
\left< r^3 \right>_{(2)} = 2.71(13) \, {\rm fm}^3 \,.
\end{equation}
We note, in passing, that this analysis excludes the 
value recently assumed in Ref.~\cite{dR2010}, which reads
\begin{equation}
\label{thirdRujula}
\left< r^3 \right>_{(2)} = 36.6(7.3) \, {\rm fm}^3 \,.
\end{equation}
The magnitude of the estimate put forward in Ref.~\cite{dR2010}
has independently been called into question~\cite{ClMi2010}.
For muonic hydrogen, the mass ratio
of the orbiting to the nuclear particle roughly is $\tfrac19$,
which is still small compared to unity.
The relative uncertainty of $\left< r^3 \right>_2$ therefore
gives a realistic indication of the
uncertainty that should be ascribed to the
$r_p^3$ term in the formula for the
Lamb shift in muonic hydrogen.

However, as stressed in Ref.~\cite{Pa1995}, the 
third Zemach moment should not be used directly for an 
evaluation of the nuclear size effect in muonic hydrogen.
In Eq.~(58) of Ref.~\cite{Pa1995}, it is shown
how to express the two-photon exchange diagram, for a finite
ratio of $m_\mu/m_p$, in terms of the proton form 
factors. Furthermore,
in Eq.~(59) of Ref.~\cite{Pa1995}, it is shown
that in leading order of $m_\mu/m_p$, the third 
Zemach moment correction is recovered. 
An evaluation of the complete two-photon 
exchange diagram is then performed for a simple
dipole model of the nuclear charge distribution,
with $G_E(-p^2) = \Lambda^4/((p^2 + \Lambda^2)^2$
and $G_M(-p^2) = \tfrac12 \, g_p \, G_E(-p^2)$.
If one assumes the dipole model, then one may relate
the third Zemach moment correction to the leading 
finite nuclear-size effect and express $\langle r^3 \rangle_{(2)}$
as being proportional to $r_p^3$, where $r_p$ is the 
root-mean-square radius of the proton. 
However, the relation of $\langle r^3 \rangle_{(2)}$ and $r_p$ is
model-dependent, as emphasized in a particularly
clear manner in Ref.~\cite{Fr1979} and 
illustrated on the Gaussian, uniform and exponential 
models for the nuclear charge distribution.

In order to obtain a realistic estimate for the 
(elastic part of the) two-photon exchange diagram,
which captures the model dependence, we thus proceed as 
follows. We first observe that for the dipole model
assumed in Ref.~\cite{Pa1995}, the full two-photon
exchange diagram yields a correction of 
$+0.018\,{\rm meV}$ to the $2P$--$2S$ Lamb shift in
muonic hydrogen. 
With a third Zemach moment based on the 
dipole model employed in Ref.~\cite{Pa1995},
which roughly amounts to $\langle r^3 \rangle_{(2)} = 2.30 \,{\rm fm}^3$
one obtains $+0.021\,{\rm meV}$
in first order in $m_\mu/m_p$ [see Eq.~(64) of Ref.~\cite{Pa1995} 
and the text following this equation],
The actual third Zemach moment is somewhat 
larger [see Eq.~\eqref{third}], 
but as shown in Ref.~\cite{Pa1995}, the reduced-mass effect
slightly reduces the nuclear-size correction of order $\alpha^5$.
The ratio of the two results
($+0.018\,{\rm meV}$ versus $+0.021\,{\rm meV}$),
which is equal to $6/7$, then gives a realistic estimate
for the reduced mass correction to the third 
Zemach moment correction in muonic hydrogen, 
beyond the factor $m_\mu/m_p$ explicitly indicated in 
Eq.~\eqref{ENS}. However, the model-independent 
value of the third Zemach 
moment should be used for the final evaluation,
as given in Eq.~\eqref{third}, and the 
correction factor $6/7$ should be applied in order to 
approximately account for the reduced-mass effects.
This gives a result of
\begin{align}
\label{EZ}
\delta E_{\rm Z} =& \; + 0.0212(12) \, {\rm meV}
\end{align}
for the $\alpha^5$ correction to the $2P$--$2S$ Lamb shift
in muonic hydrogen, which replaces the model-dependent 
term 
\begin{align}
\label{EZprime}
\delta \overline E_{\rm Z} =& \; +0.0347\,\dfrac{r_p^3}{\rm{fm}^3} \, {\rm meV}
\end{align}
used in Refs.~\cite{Pa1995,PoEtAl2010}.

%
%
\section{Reevaluation of the Proton Radius}
\label{sec4}

For convenience, we here recall that
in Ref.~\cite{PoEtAl2010}, the theoretical expression 
already given above in Eq.~\eqref{Eth_old},
\begin{equation}
\overline E_{\mathrm{th}} = 
\left( 209.9779(49) - 5.2262 \frac{r_p^2}{\mathrm{fm}^2}  + 
0.0347 \frac{r_p^3}{\mathrm{fm}^3} \right) {\rm meV},
\end{equation}
has been used 
in order to analyze the measured energy interval
$\Delta E$ defined in Eq.~\eqref{deltaEdef}.

We here add the two-body treatment of the relativistic correction to the
electronic vacuum polarization ($\delta E_a = -0.0039 \, {\rm meV}$), the
complete result for the vacuum polarization correction to the muon self energy
($\delta E_b = +0.0026 \,{\rm meV}$), and the shift due to the 
light-by-light scattering graphs ($\delta E_c = -0.0012\, {\rm meV}$),
to obtain a total theoretical shift of
\begin{equation}
\delta E = \delta E_a + \delta E_b + \delta E_c = -0.0025 \, {\rm meV}
\end{equation}
for the proton radius independent term. As the uncertainty due to 
the light-by-light scattering graphs is eliminated, the 
theoretical uncertainty of the proton-radius independent 
term shrinks from $\pm 0.0049 \, {\rm meV}$ to 
$\pm 0.0046 \, {\rm meV}$, to which we 
should add the uncertainty of $\pm 0.0012 \, {\rm meV}$ 
from Eq.~\eqref{EZ}. The model-dependent $r_p^3$ term 
given in Eq.~\eqref{EZprime}, which is valid only for a dipole
model of the nuclear charge distribution, 
is replaced by the model-independent term given in Eq.~\eqref{EZ}.
Thus, our theoretical prediction is slightly shifted from
Eq.~\eqref{Eth_old} and reads
\begin{equation}
\label{Eth_new}
E_{\rm{th}} = 
\left( 209.9974(48) - 5.2262 \frac{r_p^2}{\mathrm{fm}^2} \right) \, {\rm meV} \,.
\end{equation}
A comparison with the experimental result given in Eq.~\eqref{Eexp}
then gives a proton radius of 
\begin{equation}
\label{rp}
r_p = 0.84169(66) \, {\rm fm} \,.
\end{equation}
This value is only marginally shifted from the value given in 
Ref.~\cite{PoEtAl2010}, which reads 
$r_p = 0.84184(67) \, {\rm fm}$.
The difference of the new radius given in 
Eq.~\eqref{rp} and the CODATA recommended value 
given above in Eq.~\eqref{rpCODATA} is  
$5.0$~standard deviations and thus statistically highly significant.

%
%
\section{Conclusions}
\label{conclu}

The purpose of this paper has been twofold.  First, in Sec.~\ref{sec2}, we have
rederived all relativistic and quantum electrodynamic contributions to the $2S
\, (F\!\!=\!\! 1) \Leftrightarrow 2P_{3/2} (F\!\!=\!\!2)$ transition energy
which are relevant on the order of the reported discrepancy of theory and
experiment~\cite{PoEtAl2010}, which is $0.31 \, {\rm meV}$.  We have emphasized
that the relativistic and quantum electrodynamic corrections relevant on the
level of the discrepancy represent theoretically well-established corrections.
The QED theory of atomic bound states is generally recognized as 
a rather highly developed
theory, and the theoretical predictions for muonic hydrogen have 
been obtained as a collective effort of 
QED theorists~\cite{Pa1995,Pa1999,Pa2007,Bo2005mup,%
Bo2005mud,FaMa2003,Ma2005mup,Ma2007mup,Ma2008}. The main effects obtained 
in the cited literature references are independently being verified here.
The numerically smaller QED corrections listed in Table~II of
Ref.~\cite{Ma2005mup} beyond those treated in
Sec.~\ref{sec2} are said to be of ``higher order'' because
they are parametrically suppressed (either by higher powers in the 
mass ratio $m_\mu/m_p$ or by higher powers in the fine-structure 
constant $\alpha$). A correction of a conceivable calculational
insufficiency in one of the higher-order effects could thus only
explain the discrepancy if it leads to a 
surprising enhancement of the corresponding
correction that compensates its parametric suppression.

In Sec.~\ref{sec3}, we calculate a number of 
hitherto neglected effects which contribute to the 
$2P$--$2S$ Lamb shift in muonic hydrogen.
The most important of these probably is a fully
nonperturbative treatment of the Uehling correction to the 
bound-muon self-energy, which is performed 
using a pseudospectrum of states obtained from an
exponential grid on a lattice~\cite{SaOe1989,Jo2007}.
We also refer to Ref.~\cite{KaIvKoSh2010} for the 
results on the light-by-light scattering graphs,
and we present an updated estimate for the $\alpha^5$ correction
to the nuclear size effect, which uses a model-independent
value for the third Zemach moment~\cite{FrSi2005}.

We have already stressed that 
the theory of muonic hydrogen is given, on the level of the
discrepancy, by only few, simple relativistic and QED corrections not exceeding
the two-loop level. In some sense, this makes the task easier for theoreticians
as compared to the muon anomalous magnetic moment, where on the level of the
observed $3.4\sigma$ discrepancy of theory and experiment, a multitude of
physical effects (three-loop and beyond) contribute (see
Refs.~\cite{BeEtAl2002posmuon,BeEtAl2004negmuon,BeEtAl2006muon,HaMaNoTe2007}).
The observed discrepancy in muonic
hydrogen should thus be interpreted as a {\em large} discrepancy,
both in terms of standard deviations as well as in terms of its
absolute magnitude in energy units.

The discrepancy is all the more surprising because 
the spectroscopy of muonic bound states is an established tool for the 
determination of nuclear radii~\cite{An2004,AnEtAl2009}.
Nuclear radii determined from electron scattering and from
muonic transitions have been observed to agree on the 
percent level already in a number of experiments,
one of the first was reported in 1974 (Ref.~\cite{DuEtAl1974}).
Furthermore, the vacuum polarization contributions
to the Lamb shift (including two-loop
effects) have been verified in other muonic transitions~\cite{BeEtAl1986}.
Possibilities for a clarification of the discrepancy from a theoretical
side seem to be somewhat limited and will be explored in a following
investigation~\cite{Je2010paper2}, to the extent possible.

%
%
\section*{Acknowledgments}

The author gratefully acknowledges informative and insightful discussions with
K.~Pachucki and P.~J.~Mohr, and warm hospitality at the National Institute of
Standards and Technology in August 2010, where an important part of this work
has been performed.  Many helpful conversations with J.~Sapirstein,
S.~J.~Brodsky and U.~P.~Jentschura are also gratefully acknowledged.  This
research has been supported by a NIST Precision Measurement Grant and by the
National Science Foundation.


\rhead

\end{document}